\documentclass{webofc}
\usepackage[varg]{txfonts}   % Web of Conferences font

\usepackage{hyperref}
\usepackage{tikz}

\usepackage{caption}
\usepackage{subcaption}
\usepackage{graphicx}
\graphicspath{{./figs}}

\makeatletter 
\newcommand\semiLarge{\@setfontsize\semiLarge{13.3}{16.0}}
\makeatother

\begin{document}

\title{\semiLarge \textit{HyperTrack}: Neural Combinatorics for High Energy Physics}

\author{\firstname{Mikael} \lastname{Mieskolainen}\inst{1,2}\fnsep\thanks{\email{m.mieskolainen@imperial.ac.uk}}
}

\institute{High Energy Physics, Blackett Laboratory, Imperial College London, SW7 2AZ, United Kingdom
\and
I-X, Imperial College London, W12 0BZ, United Kingdom}

\abstract{
Combinatorial inverse problems in high energy physics span enormous algorithmic challenges. This work presents a new deep learning driven clustering algorithm that utilizes a space-time non-local trainable graph constructor, a graph neural network, and a set transformer. The model is trained with loss functions at the graph node, edge and object level, including contrastive learning and meta-supervision. The algorithm can be applied to problems such as charged particle tracking, calorimetry, pile-up discrimination, jet physics, and beyond. We showcase the effectiveness of this cutting-edge AI approach through particle tracking simulations. The code is available online.
}

\maketitle
\section{Introduction}
\label{sec:intro}

Charged particle track reconstruction is a demanding combinatorial inverse problem encountered in high energy and nuclear physics. It is also the problem which inspired the development of the very first computer vision method, the Hough transform. The future high-luminosity LHC, where each event generates $\mathcal{O}(10^4)$ charged particles and $\mathcal{O}(10^5)$ corresponding detector hits due to a mean pile-up of 200 simultaneous $pp$-collisions, calls for improved inference solutions in terms of memory, latency, and physics performance. We formulate this as a deep learned clustering problem. Earlier approaches include a discretized integral transform (Hough), template matching, recursive filtering (combinatorial Kalman filter, the de-facto method), spin-glass neural networks (Hopfield like), graph neural networks (edge prediction) and combinatorial optimization via adiabatic quantum computing.~\cite{ju2021performance}

Our \textit{HyperTrack} algorithm is built using a machine learned voxelized graph constructor and a graph neural network (GNN) which operates on the graph. Finally, a greedy or Monte Carlo random walk on the graph is seeding a set transformer (TRF), which provides the final output. This approach differs from a GNN based local or sequential tracking in clear way; we use space-time \textit{non-local} clustering, not space-time local doublet or triplet node linking per se. In terms of clustering itself, our approach differs from classical methods such as density based algorithms in a crucial way -- contrastive and meta-supervised deep learning is used to learn how to cluster.

The problem is formulated in a \textit{generic way} to address also closely related HEP problems, such as unified tracking and calorimetric object reconstruction, pile-up decomposition (hard vs soft and $N$-vertex way) and high-level physics analysis clustering problems. Furthermore, exotic particle trajectories beyond typical helix tracks in a uniform magnetic field pose no obstacle, because the underlying dynamics is learned from the training sample.

\section{Algorithm}
\label{sec:algorithm}

The full algorithm consists of (replaceable) modules, summarized in Table~\ref{tab:algotable}. Explicit description of the algorithm can be found in the public code\footnote{\href{https://github.com/mieskolainen/hypertrack}{github.com/mieskolainen/hypertrack} (MIT license)}.

\begin{table}[t]
\centering
\small
\begin{tabular}{ll}
\hline
\textbf{VD} & Input: data point vectors $\{\mathbf{x}_i\}$ \\
\hline
1) & Geometric distance search against a pre-trained voxel centroid set $\{ \mathbf{c}_k \}$ \\
2) & Look-up construction via a pre-trained $C$-matrix $\rightarrow$ an adjacency $\hat{A}$ \\
\hline
\textbf{GNN} & Input: $\{\mathbf{x}_i\}$, $\hat{A}$ \\
\hline
1) & Message-passing layer iterations $\rightarrow$ latent node embeddings $\{\mathbf{z}_i\}$ \\
2) & Evaluate 2-point MLP$_{\rho}$ $\rightarrow$ edge prediction scores between nodes $\{p_{ij}\}$ \\
\hline
\textbf{Subgraphs} & Input: $\hat{A}$, $\{p_{ij}\}$ \\
\hline
1) & Apply a global edge threshold cut $\rightarrow$ a sparsified adjacency $\tilde{A}$ \\
2) & Connected components (WCC) search $\rightarrow$ a set of disconnected subgraphs $\{\tilde{A}_k \}$ \\
3) & Sort $\tilde{A}_k$ according to their node multiplicity for batching efficiency \\
\hline
\textbf{Transformer} & Input: $\{\mathbf{x}_i\}$, $\{\mathbf{z}_i\}$, $\{\tilde{A}_k \}$, $\{p_{ij}\}$ \\
\hline
   & \textit{While} unclustered nodes in subgraphs, \textit{do} \\
1) & Find pivotal nodes per subgraph, skip nodes if max trials exceeded \\
2) & Batch (tensorize) nodes and pivotal nodes over all non-empty subgraphs \\
3) & Apply the transformer architecture in parallel to the batch tensor \\
4) & Apply a mask threshold cut to the transformer output, get passing nodes \\
5) & Collect cluster candidates, apply failure logic and remove clustered nodes \\
\hline
\end{tabular}
\caption{\textit{HyperTrack} algorithm overview.}
\label{tab:algotable}
\end{table}

\subsection{Voxel-Dynamics (VD) adjacency}

The graph neural network message passing requires a graph adjacency sparse enough, given a finite memory-latency budget, but which contains enough information for efficient reconstruction. For this we introduce a machine learned discretized estimator called \textit{Voxel-Dynamics}, which by construction has a \textit{pile-up invariant} positive (negative) edge efficiency for a given fixed voxelization. Its training consists of two parts, adaptive Voronoi voxelization of the 3D (or 4D) hit space to a number of $V$ voxels and a very sparse $V \times V$ boolean matrix $C$ describing the connectivity (dynamics) of the hits associated with voxels. It encodes the geometric "boolean bundles" of tracks passing through a voxel to another voxel, as observed during training. The voxelization is done via $K$-means algorithm using a high performance \texttt{faiss} library~\cite{johnson2019billion} in the training phase and via exact $\ell_2$-distance search during inference. Faster search algorithms are also available based on polysemous codes and quantization techniques. Training the $C$-matrix is a straightforward "forward" look-up problem achieved by iterating through the entire training sample once. The trained matrix produces a fixed ROC point, which can be controlled by adjusting parameters such as $V$ and the size of the training sample, or by optimizing the structure of the matrix afterward. This optimization could be based on the training phase integer counts per $C_{uv}$ matrix element and an additional algorithm. Typically, the matrix $C$ is very sparse for sufficiently large values of $V$.

\begin{figure}[t]
\begin{subfigure}{0.3\textwidth}

\centering
\begin{tikzpicture}[scale=0.55]
  % Define the nodes
  \node[circle, draw] (A) at (0,0)      {1};
  \node[circle, draw] (B) at (1.5,0.75) {2};
  \node[circle, draw] (C) at (3.0,1.5)  {3};
  \node[circle, draw] (D) at (4.5,2.25) {4};
  \node[circle, draw] (E) at (6.0,3.0)  {5};  
  
  % Define the curved edges
  \draw[->] (A) to[bend left] (B);
  \draw[->] (B) to[bend left] (A);
  
  \draw[->] (B) to[bend left] (C);
  \draw[->] (C) to[bend left] (B);
  
  \draw[->] (C) to[bend left] (D);
  \draw[->] (D) to[bend left] (C);
  
  \draw[->] (D) to[bend left] (E);
  \draw[->] (E) to[bend left] (D);
  
\end{tikzpicture}
\caption{\texttt{eom}-topology}
\end{subfigure}
\hfill
\begin{subfigure}{0.3\textwidth}
\centering
\begin{tikzpicture}[scale=0.55]
  % Define the nodes
  \node[circle, draw] (A) at (0,0)      {1};
  \node[circle, draw] (B) at (1.5,0.75) {2};
  \node[circle, draw] (C) at (3.0,1.5)  {3};
  \node[circle, draw] (D) at (4.5,2.25) {4};
  \node[circle, draw] (E) at (6.0,3.0)  {5};
  
  % Define the curved edges
  \draw[->] (A) to[bend left] (B);
  \draw[->] (B) to[bend left] (A);
  \draw[->] (A) to[bend left] (C);
  \draw[->] (C) to[bend left] (A);
   
  \draw[->] (B) to[bend left] (C);
  \draw[->] (C) to[bend left] (B);
  
  \draw[->] (C) to[bend left] (D);
  \draw[->] (D) to[bend left] (C);
  \draw[->] (D) to[bend left] (B);
  \draw[->] (B) to[bend left] (D);
  
  \draw[->] (D) to[bend left] (E);
  \draw[->] (E) to[bend left] (D);
  \draw[->] (C) to[bend left] (E);
  \draw[->] (E) to[bend left] (C);
\end{tikzpicture}
\caption{\texttt{cricket}-topology}
\end{subfigure}
\hfill
\begin{subfigure}{0.3\textwidth}
\centering
\begin{tikzpicture}[scale=0.55]
  % Define the nodes
  \node[circle, draw] (A) at (0,0)      {1};
  \node[circle, draw] (B) at (1.5,0.75) {2};
  \node[circle, draw] (C) at (3.0,1.5)  {3};
  \node[circle, draw] (D) at (4.5,2.25) {4};
  \node[circle, draw] (E) at (6.0,3.0)  {5};
  
  % Define the curved edges
  \draw[->] (A) to[bend left] (B);
  \draw[->] (B) to[bend left] (A);
  \draw[->] (A) to[bend left] (C);
  \draw[->] (C) to[bend left] (A);
  \draw[->] (A) to[bend left] (D);
  \draw[->] (D) to[bend left] (A);  
  
  \draw[->] (B) to[bend left] (C);
  \draw[->] (C) to[bend left] (B);
  
  \draw[->] (C) to[bend left] (D);
  \draw[->] (D) to[bend left] (C);
  \draw[->] (D) to[bend left] (B);
  \draw[->] (B) to[bend left] (D);
  
  \draw[->] (D) to[bend left] (E);
  \draw[->] (E) to[bend left] (D);
  \draw[->] (C) to[bend left] (E);
  \draw[->] (E) to[bend left] (C);
  \draw[->] (E) to[bend left] (A);
  \draw[->] (A) to[bend left] (E);
  \draw[->] (E) to[bend left] (B);
  \draw[->] (B) to[bend left] (E);  
  
\end{tikzpicture}
\caption{\texttt{hyper}-topology}
\end{subfigure}
\caption{Different graph adjacency target topologies: \texttt{eom} $\subset$ \texttt{cricket} $\subset$ \texttt{hyper} for a single track (object), where its nodes $1-5$ for \texttt{eom} and \texttt{cricket} are either time-ordered or ordered according to the minimal spanning tree. The \texttt{hyper}-topology is fully space-time order invariant. Self-edges are not visualized but are included.}
\label{fig:topologies}
\end{figure}
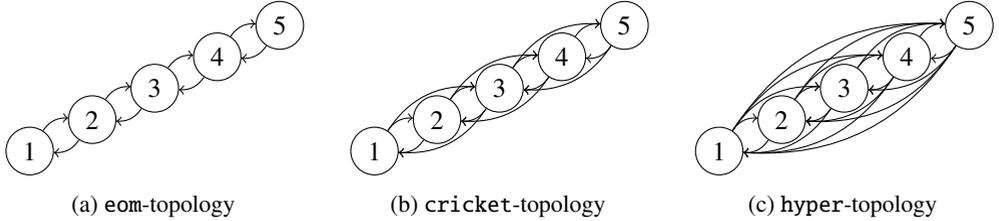

For the ground truth target adjacency, options such as equations of motions based \texttt{eom} or its multi-hop extension, \texttt{cricket}, can be feasible choices e.g. for sequential tracking. However, a space-time non-local \texttt{hyper} connectivity is chosen here because tracks are highly non-local objects, also because it provides natural "topological protection" against missing nodes and noise. The topologies are illustrated in Fig.~\ref{fig:topologies}. The idea behind \texttt{hyper} connectivity is overcompleteness. All possible edges are spanned between all the nodes of the hypothetical cluster object. A task for the neural part is to do reduction of the fake edges. Including self-edges in the graph allows the GNN message passing to discriminate between physical and non-cluster associated (noise) hits.

The inference phase can be fully parallelized. For each hit index $i$, the corresponding voxel index $u$ is searched using $\ell_2$-distance between the hit coordinate and the voxel centroid, and resulting hits per voxel are stored. Then, if $C_{uv} = 1$ for a pair of non-empty voxels $(u,v)$, the associated real space hits $(i,j)$ will obtain an adjacency value of $\hat{A}_{ij} = 1$. The sparsity structure mitigates the apparently high worst-case time complexity of nested look-ups. The VD estimator has a special formal property when applied to the tracking problem; in the limit $V \rightarrow \infty$ and infinite training statistics, the tracks could be reconstructed near perfectly. With a finite voxelization and training sample, fake edges are being generated and some real edges are lost. A key observable to consider is the mean node degree $\langle d \rangle$ in the resulting graph. In the tracking problem, the scaling law is approximately $\langle d \rangle \propto \langle \mu \rangle / V$. In essence, to control the scaling of GNN message passing complexity with an increased number of graph nodes caused by increasing the mean pile-up $\langle \mu \rangle$, it is necessary to increase the voxel count.

\subsection{Graph Neural Network architecture}

For the GNN, we use an extension of the EdgeConv~\cite{wang2019dynamic}, which we call \textit{SuperEdgeConv}. The $k$-th message passing iteration is
\begin{align}
\label{eq:gnn_mp}
\mathbf{m}_i^{(k)} &= \bigoplus_{j \in \mathcal{N}_i} \text{MLP}_\psi^{(k)} \left( \text{cat} \left[ \mathbf{h}_i^{(k)}, \mathbf{h}_i^{(k)} \odot \mathbf{h}_j^{(k)}, \mathbf{h}_j^{(k)} - \mathbf{h}_i^{(k)}, \mathbf{e}_{ij}^{(k)} \right] \right), \\
\label{eq:gnn_residual}
\mathbf{h}_{i}^{(k+1)} &\leftarrow \text{MLP}_\phi^{(k)} \left( \text{cat}\left[ \mathbf{h}_i^{(k)}, \mathbf{m}_i^{(k)} \right] \right) + \mathbf{h}_{i}^{(k)}, \;\; \text{and} \;\; \mathbf{e}_{ij}^{(k)} = \text{cat} \left[ \langle \mathbf{h}_i^{(k)}, \mathbf{h}_j^{(k)} \rangle, \|\mathbf{h}_i^{(k)} - \mathbf{h}_j^{(k)}\|, \Delta_{ij} \right].
\end{align}
The $\odot$ is an element wise (Kronecker) product, explicit edge features $e_{ij}$ include the dot product and $\ell_2$-distance, $\Delta_{ij} = (d_i - d_j) / \langle d \rangle$ is the normalized node (vertex) degree difference and the input for the first layer is $\mathbf{h}_i^{(0)} \equiv \mathbf{x}_i$. The number of message passing layers bounds the horizon of the \textit{receptive field} and $N$ iterations can see $N+1$ hops along the graph. For the track reconstruction approximately at least 3--4 layers are needed and we use 5 in the experiments. All MLP models use \texttt{silu} activation function with batch norm layers included and after the first layer, residual connections $(+ \mathbf{h}_i)$ are included in Eq.~\ref{eq:gnn_residual} to improve the gradient flow. The choice of the message passing neighbourhood $\mathcal{N}_i$ aggregator $\oplus$, typically permutation equivariant, can be critical. We choose here the \texttt{mean} aggregator. However, layer wise mixed aggregators are possible, such as using an adaptive transformer based aggregator for the last layer due to the costly $O(n^2)$ self-attention matrix operations. The choice of the aggregator can be shown theoretically to be one of the representation power bottlenecks (cf. Weisfeiler-Lehman graph isomorphism test), depending on the problem graph structure.\footnote{As an alternative design, we include an $E(N)$ group-equivariant architecture, which achieves comparable accuracy. A Lorentz group variant can also be easily obtained, especially for high-level analysis problems. An interesting open problem involves designing a GNN model that has explicit pile-up invariance properties, beyond training or conditionalizing the model under different luminosity conditions.}

Then after the message passing, GNN layer outputs are fused to obtain for each $i$-th node a latent representation
\begin{equation}
\mathbf{z}_i = \text{MLP}_\gamma \left( \text{cat} \left[ \mathbf{h}_i^{(1)}, \mathbf{h}_i^{(2)}, \dots, \mathbf{h}_i^{(K)} \right] \right),
\end{equation}
which in parallel combines representations at different depths of message passing. The embedding dimensions of $\mathbf{z}$ and $\mathbf{h}$ are important for handling the combinatorial complexity, and $\dim(\mathbf{z}) \sim \dim(\mathbf{h}) \sim 64$ seems a reasonable compromise between computational resources and representation power. Finally, the edge (2-point) scores are predicted using
\begin{equation}
p_{ij} = \sigma(l_{ij}), \;\; \text{where} \;\; l_{ij} = \text{MLP}_\rho \left( \mathbf{z}_i \odot \mathbf{z}_j \right),
\end{equation}
which is invariant under $i \leftrightarrow j$ permutation and $\sigma$ is the sigmoid function. The self-edges $i=j$ allow this function to learn to discriminate between noise and real hits.

\subsection{Subgraphs and pivotal diffusion search}

A global edge threshold cut is applied to the GNN edge scores $p_{ij} > c_e$ and a weakly connected component (WCC) graph search is done. This gives us a set of \textit{subgraphs}. The optimal case would be a "mass gap" situation between positive and negative edges, i.e., no density overlap between their edge score distributions. With the \texttt{hyper}-target connectivity, an optimal GNN would make the transformer stage unnecessary. The cut threshold is a hyperparameter tuned according to external training metrics or by adapting the cut event-by-event, e.g., using the expected and obtained mean node degree. Operationally a low cut threshold postpones clustering for the transformer, whereas a high value makes the transformer operate as a post-filter. In many clustering problems, a clear topological clustering phase transition, from a single input graph to well-disconnected subgraphs, should occur when $c_e \simeq 0.5$.

Seeding the transformer based clustering goes as follows. Per subgraph, a random set of starting nodes is selected. This can be also ordered geometrically, for example in high-level analysis, the leading transverse momentum nodes could be chosen. Then, by taking each random node as the starting node, a greedy graph walk proceeds on the subgraph along the highest log-odd probability $\log(p_{ij}/(1-p_{ij}))$ edge direction and this is iterated -- also simultaneously filtering out self-edges and previous nodes. The path which corresponds to the maximal sum of log-odds is chosen and the corresponding nodes are promoted as \textit{pivots}. This is essentially greedy directed diffusion on the subgraph. The number pivots is a free hyperparameter, which is set to three in our experiments. Alternatively, MC random walk according to a multinomial distribution spanned by the connecting edges, can be done.

After the set of pivotal nodes is found, we re-connect a \textit{micrograph}, an inclusive fully connected graph spanned by the nodes which are connected to each pivotal node. This set of nodes and chosen pivots gives us the input for the transformer. Finally, these sets are batched (tensorized) over all the subgraphs to execute the transformer in parallel.

\subsection{Transformer architecture}

A dot-production full attention transformer without positional encoding is used. This model is naturally permutation equivariant, also known as a set transformer~\cite{lee2019set}. The core function of the transformer is a scaled dot-product softmax attention function
\begin{equation}
\text{Att}(Q,K,V) = \text{softmax} \left(QK^T / \sqrt{d} \right) V,
\end{equation}
where $d$ is the input dimensionality. Queries $Q$, keys $K$ and values $V$ are named after a weak analogy with database models. These are matrices, with data vectors as rows. In most applications $V$ is set equal to $K$, which is the case also here. Batched utilization of the transformer requires boolean tensor masking applied to $QK^T$.

The actual multihead attention transformer is
\begin{align}
T &= \text{cat}[A_1, A_2, \dots, A_h] W_O, \;\; \text{where} \;\; A_j = \text{Att}(QW_j^Q, KW_j^K, VW_j^V),
\end{align}
where $W_j^Q, W_j^K, W_j^V$ and $W_O$ are learnable matrices and $j$ runs over the number of heads $h$. The total number of model parameters is the same as with a single head model, but the vectors are subspace wise split for each attention head computation separately to increase representation power, then finally concatenated and projected with a matrix $W_O$. Finally, residual connections, learnable layer normalizations and an MLP are applied
\begin{align}
H &\leftarrow \text{LayerNorm}^{(1)}(T + Q) \\
H &\leftarrow \text{LayerNorm}^{(2)}(\text{MLP}_T(H) + H).
\end{align}
Dropout regularization unit(s) could be included after the layer normalization. This whole chain of operations is denoted with $\text{MAB}(Q,K)$ operator~\cite{lee2019set} and a self-attention operator follows as $\text{SAB}(X) \equiv \text{MAB}(X,X)$. Adaptive pooling can be done via $\text{MAB}(S,X)$, where $S$ is a set of learnable vectors, but this is not used in the architecture described here.

Based on the obtained performance and complexity of different options, the developed model is as follows
\begin{align}
\textbf{Encoder:} \; &  G_{(\text{pivots})} = \text{MLP}_E(Z_{(\text{pivots})}) \\
\textbf{Decoder:} \; & D = \text{SAB}_D^{stack}(\text{MAB}_D(Q=G, K=G_{\text{pivots}})) \\
\textbf{Node score:} \; & M = \sigma( \text{MLP}_M(D)),
\end{align}
where $Z = \text{cat}[\{\mathbf{z}_i\}, \{ \mathbf{x}_i\}]$. The idea of the encoding stage is to construct an input embedding $G$ which combines both raw data and GNN processed representation to obtain end-to-end optimization. In the decoder cross-attention, the key set $G_{\text{pivots}}$ steer the attention towards a specific cluster if the query set $G$ contains several clusters. That is, the pivots explicitly break the "selection symmetry" instead of relying on a fully spontaneous breaking. Repeating the self-attention over a \textit{stack} of iterations allows to model higher order correlations beyond pairwise (up to $N$-point) and we use 4 layers. The node mask predictor $\text{MLP}_M$ provides a scalar score per node to belong to the cluster. Finally, a hard cut $M > c_n$ is applied with $c_n \simeq 0.5$, which can be optimized via gradient descent using a sigmoid based soft relaxation of the heaviside step function or adapted per batch using Fisher's variance criterion.

\subsection{Loss functions}

A binary Focal loss is used as the edge prediction loss, which is a prediction distribution entropy regularized version of the binary cross-entropy
\begin{equation}
\mathcal{L}_e = - \frac{1}{\sum_i w_i} \sum_{n=1}^E w_n \left[ y_{n}(1-p_{n})^{\gamma} \log (p_{n}) + (1-y_{n}) p_{n}^{\gamma} \log (1 - p_{n}) \right],
\end{equation}
where $E$ is the total number of edges, $p_{n} \in [0,1]$ is the edge score and $y_{n} \in \{0,1\}$ is the edge label. We use a regularization parameter $\gamma=1$, emphasizing the harder to classify edges more than $\gamma=0$ (BCE) and resulting in more well disconnected subgraphs. Integrated inverse positive-negative edge balance weights $w_n$ are used to reweight the loss.

To target the \textit{clustering goal} explicitly over $\mathcal{C}$ true clusters, a contrastive multi-object edge loss with multiple positive and negative edges is built, inspired by the $N$-pair loss~\cite{NIPS2016_6b180037} as
\begin{equation}
\mathcal{L}_c = - \frac{1}{\langle n \rangle} \frac{1}{\sum_i \omega_i} \sum_{k=1}^{\mathcal{C}} \omega_k \frac{1}{E_k^+} \sum_{j=1}^{E_k^+} \log \frac{\exp(s^+_{kj} / \tau)}{\exp(s_{kj}^+ / \tau) + \sum_{n=1}^{E_k^-} \exp(s_{kn}^- / \tau)}.
\end{equation}
The score sets $|\{s^+\}| = E^+$ and $|\{s^-\}|=E^-$ for the positive (negative) edges are the GNN edge logits $l_{ij}$ passed through a hyperbolic tangent, with self-edges excluded. A problem specific cluster weights are denoted with $\omega_k$, which are by default $\omega_k \equiv 1$. The scale normalization is $\langle n \rangle$, the mean of true cluster node multiplicities and the critical hyperparameter $\tau$ controls the dispersion of latent representations, set here to $\tau=0.3$. Including only edges that exceed a threshold, $p_{ij} > 10^{-2}$, regulates the model towards purity. This loss is computationally intensive and it can be practical to include only a subset of true clusters per event.

The transformer node mask score loss is
\begin{equation}
\mathcal{L}_n = - \frac{1}{K} \sum_{k=1}^K \frac{1}{N_k}\sum_{j=1}^{N_k} \left[ y_{kj}(1-m_{kj})^{\gamma} \log (m_{kj}) + (1-y_{kj}) m_{kj}^{\gamma} \log (1 - m_{kj}) \right],
\end{equation}
where $m_{kj}$ is the transformer mask predictor score for the $j$-th node of the $k$-th estimated cluster, $N_k$ is the number of nodes in $Z$ (transformer input) and the number of estimated clusters is $K$. \textit{Meta-supervision} is applied to construct targets $y_{kj} \in \{0, 1\}$. In our scheme, the dominant ground truth cluster label within the pivotal nodes defines the true class.

The node set loss is used to optimize the final clustering result as
\begin{equation}
\mathcal{L}_s = - \frac{1}{N} \sum_{k=1}^K \left( \sum_{j \in \Upsilon_k \cap Y_k} m_{kj} - \sum_{j \in \Upsilon_k - Y_k} m_{kj} \right),
\end{equation}
where the first term is an intersection between the estimate $\Upsilon_k$ and ground truth $Y_k$ node sets, targeting efficiency and the second term is a set difference, targeting purity. The set $\Upsilon_k$ includes nodes which passed the final cluster mask threshold cut and the cluster ground truth is given by the meta-supervision. $N$ is the total number of data points.

Finally, the total hybrid loss to minimize is $\mathcal{L} = \sum_i \beta_i \mathcal{L}_i$, where $\beta_i$ are relative strength hyperparameters and one of them can be set to a constant. We use unoptimized values $\beta_e = \beta_n = \beta_s = 0.2$ and $\beta_c = 1.0$. An open problem is to find a way to balance them automatically.

\section{Experiments}
\label{sec:experiments}

As an extensive proof-of-concept, we study the track reconstruction problem. Only the 3D coordinates of the detectors hits $\mathbf{x}=[x,y,z]$ are used as an input, without any pre-transformation. The charge deposit amplitudes could improve the performance, similarly time-domain information could be incorporated. We use the Kaggle TrackML~\cite{amrouche2020tracking} dataset of simulated $pp$-collisions at $\sqrt{s}=14$ TeV made using Pythia 8 and ACTS fast detector simulation over $\eta \in [-4,4]$. The Poisson average pile-up is $\langle \mu \rangle = 200$, resulting in around $10^5$ detector hits (nodes) and $10^4$ tracks (clusters). Two simplifications are applied: the track density is decreased to simulate pile-up of $\langle \mu \rangle = (2,20,60)$ and the noise hit fraction (non-associated hits) is reduced from 15\% to 5\%. No additional kinematic cuts are made and Kaggle competition hit weights are used in the contrastive loss and evaluating the double majority score (DMS) values. The dataset contains around $9\times10^3$ events in total, which we split into VD training (25\%), neural training (60\%) and validation-evaluation (15\%).

Our hardware includes an NVIDIA V100 with 32 GB of VRAM and an Intel Xeon Gold 6230 with 20 physical cores and 86 GB of RAM. The most important libraries in use are \texttt{torch}, \texttt{torch-geometric}~\cite{fey2019fast} and \texttt{faiss}~\cite{johnson2019billion}. The code implementation is high-level, but most pure Python functions have been JIT-compiled. The gradient search uses AdamW optimizer with a base learning rate of $\lambda = 5 \times 10^{-4}$ with gradients updated after every event (batch size 1) and a cosine scheduler ($\lambda / 10$) with warm restarts every $10^4$ iterations, and we employ simple transfer learning by continuing the high pile-up training from the low pile-up model progressively. The transformer training is activated end-to-end once the GNN is nearly converged. The VD consists of $3 \, \times \, V \approx 1.6$ M real and $V^2 \approx 275$ G (0.4 G non-zero) boolean parameters, when $V = 2^{19}$, which generates around $2$ M edges for $\langle \mu \rangle = 60$. Inference demands considerably less memory than training. Approximately only 8 GB of VRAM and 10 GB of RAM are sufficient for $\langle \mu \rangle = 60$, because we utilize here high voxel count $V$ and maintain modest depth and width for the neural model, which consists of only 0.3 M parameters. The inference is technically feasible even for $\langle \mu \rangle = 200$. Memory constraints (for training) can be extended by using reduced precision (e.g., BFloat16), model pruning, memory efficient optimizers, piece-wise training and recent GPUs with 144 GB of VRAM. The training was executed for a total of approximately $0.5 \times 10^6$ iterations. 

\begin{figure}[t]
\centering
\begin{subfigure}{.5\textwidth}
  \centering
  \includegraphics[width=1.0\linewidth]{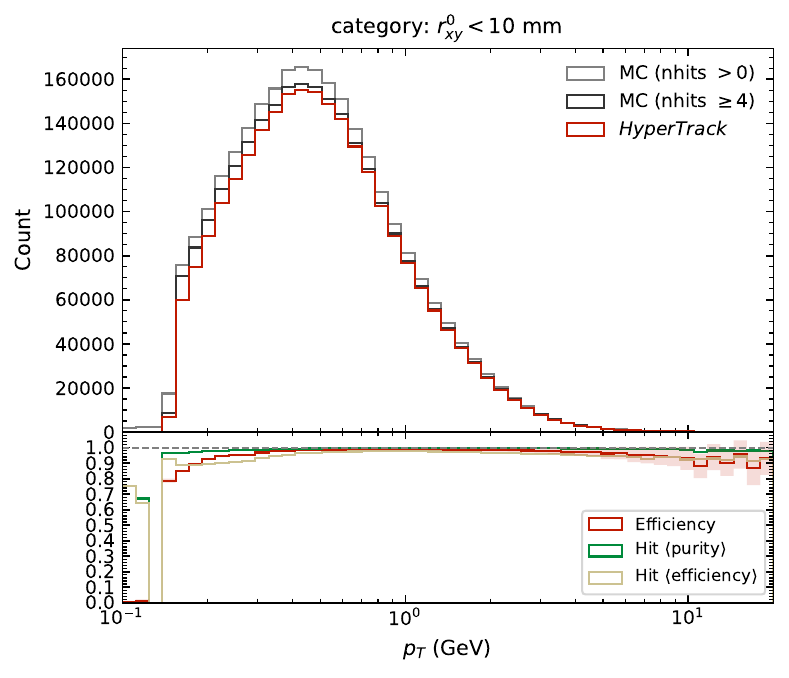}
  %\caption{A subfigure}
  %\label{fig:sub1}
\end{subfigure}%
\begin{subfigure}{.5\textwidth}
  \centering
  \includegraphics[width=1.0\linewidth]{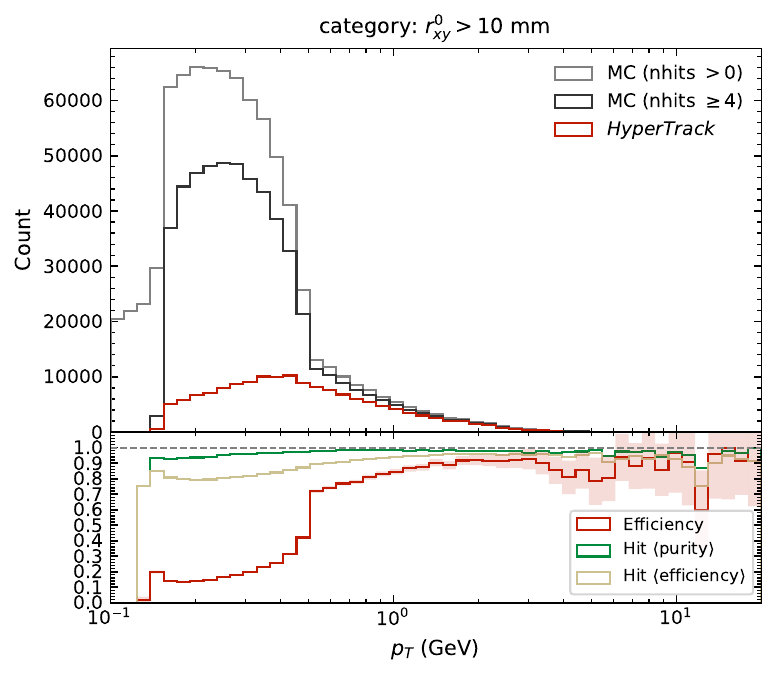}
  %\caption{A subfigure}
  %\label{fig:sub2}
\end{subfigure}
\caption{Clustering performance for $\langle \mu \rangle$ = 60. The transverse momentum for prompt (< 10 mm) and displaced vertex tracks (> 10 mm), requiring at least 4 hits. The ratios show efficiency of $\text{DMS}_{\geq 4}$ matched tracks and the hit set purity \& efficiency of the matched tracks.}
\label{fig:performance}
\end{figure}

\subsection{Results}

The inference accuracy and latency $t$ (in sec) scaling results are summarized in Table~\ref{tab:results}. The true and false positive edge efficiencies (TPR, FPR) of the VD stage are denoted as $\text{VD}_{\epsilon_{+(-)}}$ and evaluated with respect to the \texttt{hyper}-topology ground truth. The GNN edge efficiencies are with respect to the GNN input adjacency list ground truth labels and the GNN working point is at the edge cut threshold $c_{e} = 0.55$ for all scenarios. In terms of latencies, both GNN and transformer networks utilize the GPU. The rest of the computations are performed on the CPU, including the transformer input processing (subgraph logistics, pivotal search), which is currently unoptimized and accounts for over 95\% of its total processing time $t_{TRF}$.

Figure~\ref{fig:performance} shows the physics performance for $\langle \mu \rangle = 60$. The results demonstrate excellent clustering of prompt tracks, while the challenges arise with low transverse momentum tracks that exhibit high vertex displacement within the central pseudorapidity region (loopers), originating from long-lived particle decays or secondary interactions with materials such as gamma conversions $\gamma \rightarrow e^{+}e^{-}$. To improve the true positive edge efficiency VD$_{\epsilon_+}$, a larger training sample is needed for the $C$-matrix, especially with $V=2^{19}$. To prioritize latency over accuracy, it is possible to utilize only the edge cut and WCC search after VD+GNN, especially for lower pile-up scenarios. Combining VD+GNN+cut with classic density based clustering is also an option, based on sparse metric distances $(1 - p_{ij})$ obtained via GNN scores, which is available in the code using \texttt{[h]dbscan}. By design, the transformer is the method of choice for meeting the highest efficiency and purity requirements.

\begin{table}[t]
\centering
\tiny
\begin{tabular}{c|ll|llc|ccc|llll}
\hline
$\langle \mu \rangle$ & $\text{VD}_{\epsilon_+}$ & $\text{VD}_{\epsilon_-}$ & $\text{GNN}_{\epsilon_+}$ & $\text{GNN}_{\epsilon_-}$ & AUC & DMS$_{\geq 4}$ & $\mathcal{E}$ & $\mathcal{P}$ & $t_{VD}$ & $t_{GNN}$ & $t_{WCC}$ & $t_{TRF}$ \\
\hline
$2$  & 0.97(1) & $0.020(2)$ & 0.992(5) & $0.003(1)$ & 0.9998(3) & 0.95(2) & 0.96(2) & 0.99(2) & 0.0054(8) & 0.0075(5) & 0.014(2) & 0.08(1) \\
$20$ & 0.92(1) & $0.0040(2)$ & 0.983(3) & $0.0037(4)$ & 0.9995(2) & 0.89(1) & 0.89(1) & 0.94(1) & 0.13(2) & 0.0089(5) & 0.12(2) & 0.64(9) \\
$60$ & 0.86(2) & $0.00170(4)$ & 0.959(6) & $0.0044(2)$ & 0.9985(3) & 0.83(2) & 0.85(1) & 0.93(2) & 0.9(2) & 0.010(1) & 0.58(7) & 1.8(3) \\
\hline
\end{tabular}
\caption{\small Inference mean values (and std) for three pile-up scenarios using $10^3$ events, taking into account \textit{both} primary and secondary particles. The DMS$_{\geq 4}$ represents the double majority score, $\mathcal{E} = |\, \text{estimated} \cap \text{matched} \,| \, / \, |\, \text{simulated} \wedge n_{hits} \geq 4 \,|$ is the clustering efficiency and $ \mathcal{P} = |\, \text{estimated} \cap \text{matched} \,| \, / \, |\, \text{estimated} \,|$ is the clustering purity. The VD voxel counts are $V=(2^{16}, 2^{18}, 2^{19})$ per scenario. Note that the VD edge efficiencies are pile-up invariant when $V$ is fixed.}
\label{tab:results}
\end{table}

\section{Conclusions}
\label{sec:conclusions}

We introduced a new generic AI-driven clustering algorithm called \textit{HyperTrack} and demonstrated its strong performance in charged particle tracking simulations, with pile-up means up to \( \langle \mu \rangle = 60 \), which corresponds to $\sim 3\,000$ track clusters per event. This fully trainable and space-time non-local algorithm also allows for simultaneous learning of cluster sub-structure mechanics and targeting physics-constrained loss functions, capabilities that are beyond the reach of classical algorithms. In the future, we can end-to-end integrate the cluster (track) parameter regression directly into the transformer output, investigate hierarchical or recursive and $N$-point (higher rank) VD, address other learnable clustering problems and explore quantum computing options such as accelerating the VD with Grover's search. Additionally, adaptive sparsification of the cluster ground truth target topology, is an intriguing direction.
\\
\vspace{-0.5em}

\noindent \textbf{Acknowledgements}: Thanks to Alex Tapper, Liv Vage and Simon Williams for discussions. The author is supported by the Schmidt AI in Science fellowship of Schmidt Futures and I-X.

\vspace{-0.65em}
% BibTeX or Biber users please use (the style is already called in the class, ensure that the "woc.bst" style is in your local directory)
\bibliography{bibtex}

\begin{thebibliography}{7}

\bibitem{ju2021performance}
X.~Ju, D.~Murnane, P.~Calafiura et~al., EPJC \textbf{81}, 1 (2021)

\bibitem{johnson2019billion}
J.~Johnson, M.~Douze, H.~J{\'e}gou, IEEE Transactions on Big Data \textbf{7},
  535 (2019)

\bibitem{wang2019dynamic}
Y.~Wang, Y.~Sun, Z.~Liu et~al., TOG \textbf{38}, 1 (2019)

\bibitem{lee2019set}
J.~Lee, Y.~Lee, J.~Kim et~al., in \emph{ICML} (2019), pp. 3744--3753

\bibitem{NIPS2016_6b180037}
K.~Sohn, in \emph{NeurIPS} (2016), Vol.~29

\bibitem{amrouche2020tracking}
S.~Amrouche, L.~Basara, P.~Calafiura et~al., \emph{The tracking machine
  learning challenge: accuracy phase} (Springer, 2020)

\bibitem{fey2019fast}
M.~Fey, J.E. Lenssen, arXiv:1903.02428  (2019)

\end{thebibliography}

\end{document}